\documentclass[traditabstract]{aa}  

\usepackage{graphicx}
\usepackage{txfonts}
\usepackage{hyperref}
\usepackage[normalem]{ulem}
\usepackage{xcolor}

\begin{document}

\title{X-ray properties of RR Lyrae and Cepheid variables from eROSITA}

   \author{Krystian I\l{}kiewicz
          \inst{1}\thanks{ilkiewicz@camk.edu.pl}
          \and
          Henryka Netzel\inst{1}\thanks{henia@netzel.pl}
          }

   \institute{Nicolaus Copernicus Astronomical Centre, Polish Academy of Sciences, Bartycka 18, PL-00-716 Warszawa, Poland
             }

   \date{}

  \abstract {
We present a search for X-ray counterparts to RR Lyrae and Cepheid variables using data from the first eROSITA all-sky survey. We identify seven RR~Lyrae and eight Cepheid variables with positional matches to X-ray sources. While most Cepheid associations appear reliable, the RR Lyrae matches are predominantly spurious. Only one source, OGLE-BLG-RRLYR-00252, appears to be a plausible RR Lyrae detection, potentially representing the first observational evidence of X-ray emission from a star of this type. Its inferred luminosity suggests that RR Lyrae stars are intrinsically at least two orders of magnitude fainter in X-rays than the brightest Cepheids. We also observe a tentative increase in X-ray luminosity with pulsation period among classical Cepheids, and higher luminosities in type II Cepheids at comparable periods. These trends may reflect intrinsic differences in atmospheric structure and shock efficiency, offering new insight into the mechanisms driving high-energy emission in pulsating stars.

}

   \maketitle

\section{Introduction}\label{sec:introduction}

RR Lyrae stars are low-mass, Population II pulsating variables located on the horizontal branch, where they intersect the classical instability strip. They primarily oscillate in the radial fundamental mode (RRab type RR Lyrae), the first overtone (RRc), or both simultaneously (RRd). In rare cases they pulsate in the fundamental and second overtone modes simultaneously \citep{benko2010, moskalik2013}, and a few have been found to exhibit triple-mode pulsations involving three radial modes \citep{molnar2012, soszynski2014, jurcsik2015}.

Classical Cepheids (hereafter DCEPs) are intermediate-mass, Population I stars that cross the instability strip during their evolution across the Hertzsprung gap or along a core-helium-burning blue loop. Most DCEPs pulsate in a single radial mode, either the fundamental or first overtone, although double-mode and triple-mode Cepheids are also observed \citep[e.g.][]{soszynski2015}.

Type II Cepheids (T2) are low-mass, Population II pulsators that follow a distinct period–luminosity relation from that of DCEPs \citep[see Fig.~1 in][]{soszynski2008}. They are subdivided into three classes based on their pulsation period: BL Herculis (BL Her), W Virginis (W Vir), and RV Tauri (RV Tau), arranged in order of increasing period. These subclasses correspond to different evolutionary stages: from post-horizontal-branch evolution towards the asymptotic giant branch (BL Her), through helium-shell flashes (W Vir), to post-asymptotic giant branch evolution en route to the white dwarf cooling sequence (RV Tau; \citealt{gingold1976}).

Anomalous Cepheids (aCep or BL Boo), with masses in the range $1 - 2\,{\rm M}_\odot$, occupy an intermediate position on the period–luminosity diagram between DCEPs and Type II Cepheids \citep[see Fig.~1 in][]{soszynski2008}. Both T2 and aCeps exhibit pulsation characteristics akin to those of RR Lyrae stars and DCEPs.

All of these classical pulsators obey period–luminosity relations, making them valuable standard candles for distance measurements. In addition to their primary radial pulsations, many display secondary phenomena such as amplitude and phase modulations or additional periodicities (see \citealt{smolec2021_review} and references therein). In RR Lyrae stars, such quasi-periodic modulations are collectively referred to as the Blazhko effect, and their origin remains unresolved \citep{blazhko}. Blazhko-like behaviour has also been observed in DCEPs and Type II Cepheids \citep{smolec2017, smolec2018}. Moreover, additional periodicities — some likely caused by low-amplitude non-radial modes — are a common feature across classical pulsators. These have been detected in RR Lyrae stars, DCEPs \citep[e.g.][and references therein]{netzel_review}, and one anomalous Cepheid \citep{2021ApJS..253...11P}.

High-amplitude pulsations can generate shock waves that propagate through the stellar atmosphere. These shocks manifest observationally as phase-dependent helium and hydrogen emission lines or as spectral line doubling \citep[e.g.][]{preston2009, gillet.fokin2014, hocde2020, anderson2024}.

In recent years, X-ray emission has been detected in several DCEPs --- $\delta$~Cep, $\beta$~Dor, and Polaris --- using observations from \textit{XMM-Newton} and \textit{Chandra} \citep{engle2017, evans2020_xray_v473lyr, 2022ApJ...938..153E}. The observed X-ray emission is transient and strongly phase-dependent, suggesting a direct connection to pulsation-induced shock activity. Hydrodynamical models by \citet{moschou2020} demonstrate that such shocks can reproduce the observed X-ray light curves, with emission peaking at phases $0.4 -0.8$ and remaining quiescent at other phases. More recently, \citet{Fraschetti2023} conducted a detailed spectral analysis of $\delta$~Cep and identified a likely non-thermal X-ray component in addition to the shock-driven emission.

Theoretical work by \citet{bejgman.stepanov1981} postulated that RR Lyrae pulsations produce soft X-ray emission in the $80$–$100$\,\AA\ band, with luminosities of the order of $10^{30}$\,erg\,s$^{-1}$ in this range and up to $10^{32}$\,erg\,s$^{-1}$ in the $100$–$200$\,\AA\ band. Similar conclusions were drawn by \citet{andrievskii1993}. Nevertheless, no confirmed detections of X-ray emission from RR Lyrae stars have been reported to date.

In this study we searched for X-ray emission from RR Lyrae stars, DCEPs, Type II Cepheids, and anomalous Cepheids using the extended ROentgen Survey with an Imaging Telescope Array (eROSITA). In Sect.~\ref{sec:catalogue_search} we describe the methodology, the results are presented in Sect.~\ref{sec:results}, and conclusions are summarized in Sect.~\ref{sec:conclusions}.

\section{Catalogue search}\label{sec:catalogue_search}

We investigated X-ray emission from classically pulsating stars using data from eROSITA \citep[][]{2012arXiv1209.3114M,2021A&A...647A...1P}. Specifically, we utilized the first all-sky survey catalogue from eROSITA telescope array aboard the Spektrum Roentgen Gamma (SRG) satellite (\texttt{eRASS1}), which covers approximately half of the sky, corresponding to Galactic longitudes in the range $359.94423568 > l > 179.94423568$~degrees \citep{2024A&A...682A..34M}.

We compiled a catalogue of RR~Lyrae stars from the General Catalogue of Variable Stars (GCVS) version 5.1 \citep{2017ARep...61...80S}, as well as from the Optical Gravitational Lensing Experiment (OGLE) Survey \citep{2019AcA....69..321S} and stars identified in Transiting Exoplanet Survey Satellite (TESS) data \citep{2022ApJS..258....8M}. To remove duplicate entries, sources within 3~arcseconds of each other were considered identical and merged. After cross-matching and removing duplicates, the final sample included 90,327 RR~Lyrae stars. After limiting the RR~Lyrae sample to the portion of the sky covered by \texttt{eRASS1}, 43,100 stars remained.  Similarly, we assembled a catalogue of Cepheid variables from GCVS v5.1 \citep{2017ARep...61...80S}, the OGLE Survey \citep{2020AcA....70..101S}, and TESS \citep{2021ApJS..253...11P}, applying the same procedure to remove duplicate entries. The resulting sample includes 5,003 Cepheids, of which 3,084 fall within the sky coverage of \texttt{eRASS1}. This included 2,004 DCEPs, 997 type~II Cepheids and 83 anomalous Cepheids.

Various statistical methods are commonly employed to identify optical counterparts of X-ray sources \citep[e.g.][]{2018MNRAS.473.4937S}. However, in the case of RR~Lyrae stars, no robust X-ray detections have been reported to date, precluding the development of statistically grounded models tailored to their X-ray emission characteristics. Similarly, the number of known X-ray-detected Cepheids remains too small to support a comprehensive understanding of their high-energy properties. Consequently, rather than relying on probabilistic association techniques, we adopted a direct spatial cross-matching approach to identify candidate X-ray-emitting classical pulsators.

We performed an initial cross-match between the \texttt{eRASS1} catalogue and our classical pulsator sample using a 20~arcsecond matching radius. We then refined the matches by excluding cases where the separation between the pulsator and the X-ray source exceeded the positional uncertainty of the X-ray source. However, this criterion proved too restrictive, as the known X-ray-bright Cepheid \object{$\beta$~Dor} \citep{2022ApJ...938..153E} was not recovered, as the reported X-ray positional uncertainty was 0.98~arcseconds, while the separation between the optical and X-ray coordinates was 1.06~arcseconds. We attribute this discrepancy to astrometric uncertainties affecting very bright stars ($G < 6$~mag; \citealt{2018A&A...616A...2L}) and potential contamination of the eROSITA detector by optical loading. Accordingly, for stars with $G < 6$~mag, we retained matches with separations up to 5~arcseconds, even when this exceeded the formal X-ray positional uncertainty. We note that this procedure excluded one promising RR~Lyrae candidate with an X-ray detection - \object{RV Men}. This RR~Lyrae star lies $\sim$5~arcseconds from the known X-ray source \object{1RXS J061711.3-732900}, which is also detected in \texttt{eRASS1} and, to our knowledge, has no other nearby optical counterpart.

\section{Results}\label{sec:results}

As a result of our search, we have identified 16 RR~Lyrae stars whose positions matched X-ray sources in the \texttt{eRASS1} catalogue. Among these, two RR~Lyrae stars lie in the close vicinity of bright X-ray sources, and the detections are likely spurious. Specifically, \object{OGLE-GD-RRLYR-08028} is located near the bright high-mass X-ray binary \object{V821~Ara}, and \object{OGLE-BLG-RRLYR-43033} lies close to the low-mass X-ray binary \object{V1101~Sco}. In both cases, the detected X-ray emission is likely unrelated to the RR~Lyrae stars. Similarly, the X-ray source near \object{OGLE-BLG-RRLYR-44992} is likely part of an extended supernova remnant associated with \object{SN~393}. In another case, \object{V0828~CrA} is located within a globular cluster, making it unlikely that the nearby X-ray source is directly associated with the RR~Lyrae star. Among the remaining RR~Lyrae stars with X-ray counterparts, five appear to be misclassified. The classification of \object{V3166~Sgr} as an RR~Lyrae star has already been questioned by \citet{2019A&A...623A.117K}, who suggest it is more likely a young stellar object \citep{2022AJ....163...64E}. \object{V0387~Vir} is potentially a misclassified quasar, corresponding to \object{2QZJ122221.5-012438}. \object{IT~Oph} appears in several emission-line star surveys, raising doubts about its RR~Lyrae classification \citep{1964CoBos..27....1T,1977ApJS...33..459S}. Moreover, its All Sky Automated Survey for SuperNovae (ASAS-SN) light curves in the $g$ and $V$ bands show no variability characteristic of RR~Lyrae stars \citep{2014ApJ...788...48S,2023arXiv230403791H}. \object{GQ~Lib} has been reported as a long-period variable \citep{2018AJ....156..241H}, and no detailed study has confirmed its RR~Lyrae classification. Finally, \object{V1154~Ori}, originally classified as an RRc star, was shown to be non-variable by \citet{jerzykiewicz2003}. After excluding these sources, seven RR~Lyrae candidates with detected X-ray emission remain, as summarized in Table~\ref{table:rrlyr}.

\begin{table*}
\centering
\caption{RR Lyrae stars with likely X-ray counterparts in the 0.2-2.3~keV range.}
\label{table:rrlyr}
\begin{tabular}{cccccccc}

\hline\hline
Name & Type & Sep & posErr & $\sigma_{\mathrm{DET}}$ & Distance & Flux & Luminosity \\
     &      & [arcsec] & [arcsec] &     & [pc]   & [$\mathrm{mW\,m^{-2}}$] & [$\mathrm{erg\,s^{-1}}$]\\
\hline
OGLE-BLG-RRLYR-00252 & RRab & 3.07 & 3.11 & 19.5 & $168.10_{-0.53}^{+0.56}$ & $(1.07\pm 0.34) \times 10^{-13}$ & $3.62_{-1.17}^{+1.18} \times 10^{29}  $ \\
OGLE-GD-RRLYR-03705 & RRc & 3.52 & 4.71 & 9.80 & $6528_{-1746}^{+2452}$ & $(4.00\pm 1.86) \times 10^{-14}$ & $2.04_{-1.45}^{+3.62}  \times 10^{32}$ \\
OGLE-GD-RRLYR-07862 & RRab & 7.33 & 7.69 & 7.63 & $2823_{-751}^{+1417}$ & $(5.26\pm 2.36) \times 10^{-14} $ & $5.02_{-3.53}^{+11.4}\times 10^{31} $ \\
OGLE-BLG-RRLYR-05650 & RRc & 3.56 & 7.46 & 7.41 & $7188_{-2227}^{+2717}$ & $(6.70\pm 2.96) \times 10^{-14}$ & $4.14_{-3.04}^{+7.20} \times 10^{32} $ \\
OGLE-BLG-RRLYR-41100 & RRab & 2.90 & 5.71 & 6.30 & $6738_{-1210}^{+2199}$ & $(3.83\pm 2.03) \times 10^{-14} $ & $2.08_{-1.42}^{+3.52} \times 10^{32}  $ \\
V3732 Sgr & RRab & 2.24 & 5.09 & 5.95 & $7825_{-1297}^{+2015}$ & $(5.20 \pm 2.87) \times 10^{-14} $ & $3.81_{-2.62}^{+5.54} \times 10^{32}  $ \\
OGLE-BLG-RRLYR-20811 & RRab & 6.00 & 6.52 & 5.34 & $3662_{-1708}^{+3082}$ & $(4.21\pm 2.38) \times 10^{-14}  $ & $6.76_{-5.92}^{+2.91} \times 10^{31}$ \\
\hline
\end{tabular}
\tablefoot{
Sep - separation between the star and X-ray source on the sky, posErr - uncertainty of the X-ray source position, $\sigma_{\mathrm{DET}}$ - detection likelihood in 0.2-2.3~keV band.}
\end{table*}

Among the Cepheids, eleven sources are found to have X-ray counterparts in \texttt{eRASS1}.  Among them, two were present in the original OGLE Cepheids catalogue, but were later removed from the list of Cepheids (\object{OGLE-GD-T2CEP-0036} and \object{OGLE-GD-T2CEP-0118}). Moreover, \object{UW For} is an eclipsing binary and not a pulsating star \citep{2019MNRAS.486.1907J}. This leaves eight Cepheids with X-ray counterparts, including three stars with uncertain classification in the OGLE catalogue (\object{OGLE-BLG-T2CEP-1084}, \object{OGLE-GD-T2CEP-0065}, \object{and OGLE-GD-T2CEP-0007}). Our candidates are listed in Table~\ref{table:cep}. Among these, five are DCEPs and three are type~II Cepheids. Only one of these stars, \object{$\beta$~Dor}, had previously been detected in X-rays. The remaining X-ray detections, including those associated with both Cepheids and RR Lyrae stars, represent new candidate identifications.

\begin{table*}
\begin{center}
\caption{Same as Table~\ref{table:rrlyr} but for Cepheids.}
\label{table:cep}
\begin{tabular}{cccccccc}
\hline\hline
Name & Type & Sep & posErr & $\sigma_{\mathrm{DET}}$ & Distance & Flux & Luminosity \\
     &      & [arcsec] & [arcsec] &     & [pc]   & [$\mathrm{mW\,m^{-2}}$] & [$\mathrm{erg\,s^{-1}}$]\\
\hline
bet Dor\tablefootmark{*} & CEP & 1.06 & 0.98 & 1941 & $343_{-15}^{+14}$ & $(2.69 \pm 0.13) \times 10^{-13}$ & $3.78_{-0.50}^{+0.52} \times 10^{30}$ \\
l Car\tablefootmark{*} & CEP & 3.57 & 1.07 & 1104 & $515_{-29}^{+33}$ & $(6.11  \pm 0.42) \times  10^{-13}$ & $1.94_{-0.34}^{+0.41} \times 10^{31}$ \\
OGLE-BLG-T2CEP-1084 & T2 & 2.49 & 2.55 & 64.48 & $1227_{-28}^{+35}$ & $(2.40 \pm 0.47) \times 10^{-13} $ & $4.33_{-0.99}^{+1.14} \times 10^{30}$ \\
zet Gem\tablefootmark{*} & CEP & 0.78 & 3.13 & 23.1 & $326_{-23}^{+28}$ & $(8.81 \pm 3.36) \times 10^{-14}$ & $1.12_{-0.53}^{+0.71} \times 10^{30}$ \\
OGLE-GD-T2CEP-0065 & T2 & 4.26 & 4.85 & 9.16 & $3673_{-456}^{+476}$ & $(3.85 \pm 1.68) \times 10^{-14}$ & $6.21_{-3.53}^{+5.18} \times 10^{31}$ \\
AX Cir & CEP & 1.12 & 4.46 & 6.42 & $550_{-76}^{+101}$ & $(2.33 \pm 1.23) \times 10^{-14}$ & $8.42_{-5.48}^{+9.62} \times 10^{29}$ \\
MY Pup & CEP & 4.73 & 4.93 & 5.46 & $815_{-26}^{+29}$ & $(1.49  \pm 0.90) \times 10^{-14}$ & $1.18_{-0.75}^{+0.86}  \times 10^{30}$ \\
OGLE-GD-T2CEP-0007 & T2 & 4.25 & 5.49 & 5.23 & $1342_{-127}^{+160}$ & $(3.53 \pm 2.19) \times 10^{-14}$ & $7.61_{-5.24}^{+7.84} \times 10^{30}$ \\
\hline
\end{tabular}
\tablefoot{
\tablefoottext{*}{X-ray flux likely affected by optical loading.}
}
\end{center}
\end{table*}

The majority of sources have measured counts insufficient for spectral analysis. Consequently, we adopted the X-ray flux estimates directly from the \texttt{eRASS1} catalogue, where fluxes are calculated assuming an absorbed power-law model with a photon index of 2.0 and a typical Galactic absorbing column density \citep{2022A&A...661A...1B,2024A&A...682A..34M}. This model is appropriate for active galactic nuclei but may not accurately represent the X-ray emission from pulsating stars \citep[e.g.][]{Fraschetti2023}. As a result, the reported fluxes may systematically deviate from the true values. To ensure uniformity across the sample and facilitate a consistent comparison of the X-ray properties of Cepheids and RR~Lyrae stars, we used the catalogue-provided fluxes even for sources with sufficient counts for spectral modelling. This approach avoids introducing biases that could arise from selectively applying different models.

The potential impact of the assumed spectral model is evident in several cases. For example, the DCEP \object{$\beta$~Dor} was previously detected in X-rays with a luminosity approximately one order of magnitude lower than the value listed in Table~\ref{table:cep} \citep{2022ApJ...938..153E}. Similarly, \object{$\ell$ Car} appears two orders of magnitude brighter in the \texttt{eRASS1} catalogue compared to earlier upper limits on its X-ray luminosity \citep{2022ApJ...938..153E}. However,  both X-ray detections from these sources are likely affected by optical loading, which can artificially enhance the measured X-ray fluxes, leading to overestimations. For this reason, these flux values should be treated with caution and are best used for internal comparisons within our sample, rather than for direct comparison with measurements obtained using other instruments or analysis techniques.

\begin{figure}
   \centering
   \includegraphics[width=\columnwidth]{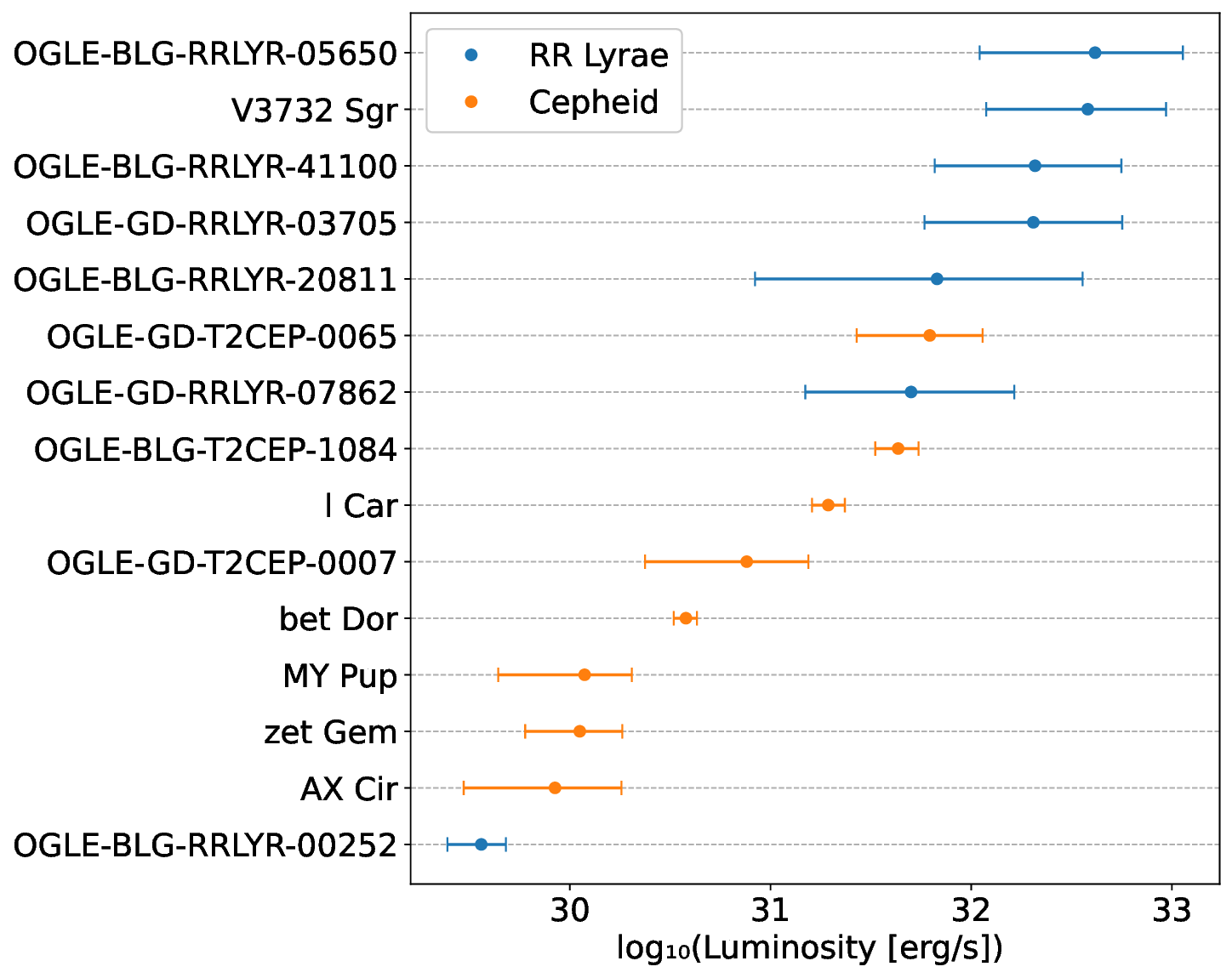}
      \caption{Sorted X-ray luminosities of RR Lyrae stars (blue) and Cepheids (orange) detected with eROSITA. Horizontal error bars represent uncertainties arising from flux measurement errors and distance uncertainties. }
         \label{fig:luminosity_errorbars}
   \end{figure}

\begin{figure*}
   \centering
   \includegraphics[width=0.99\columnwidth]{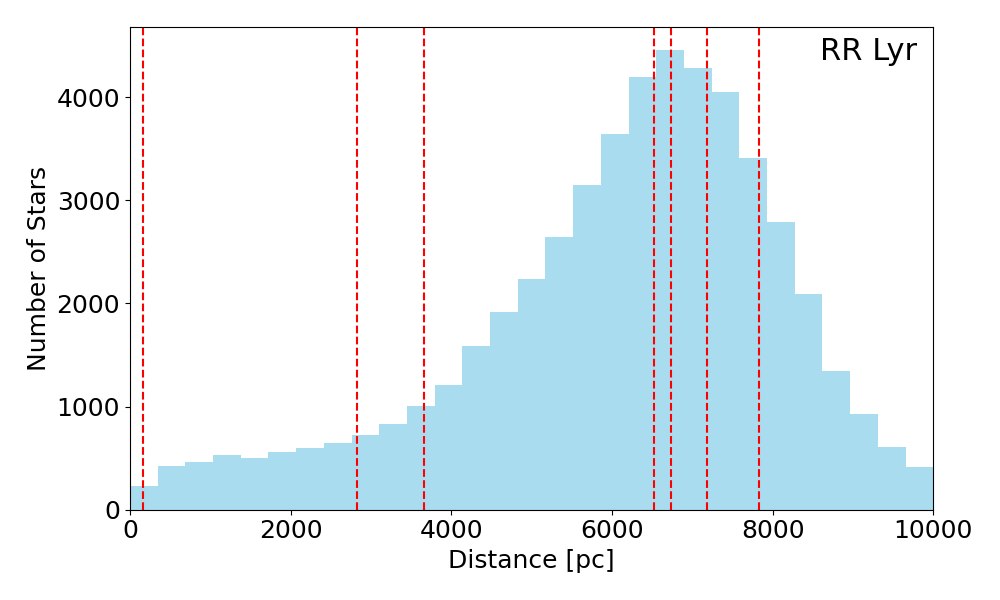}
   \includegraphics[width=0.99\columnwidth]{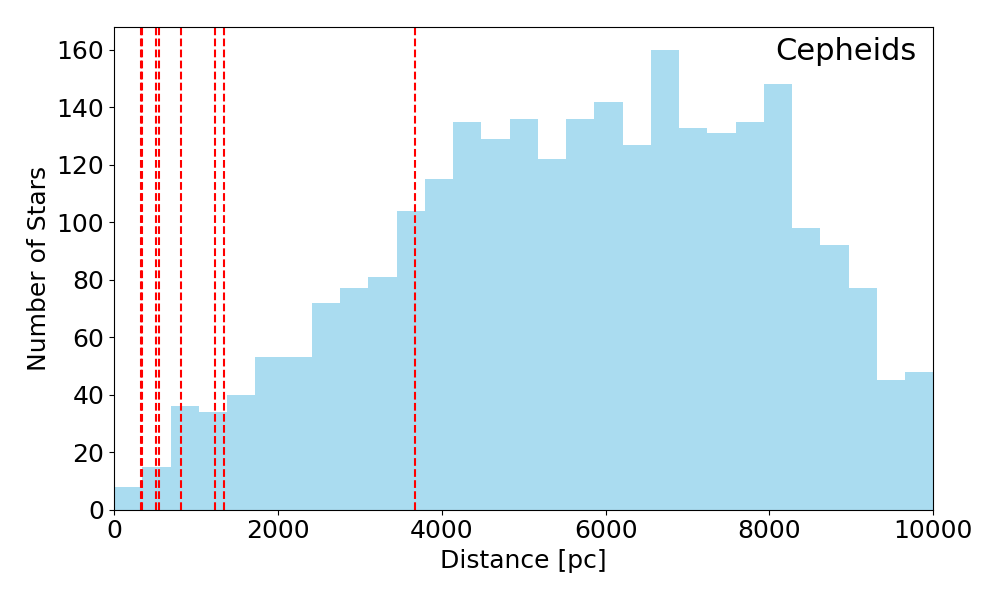}
      \caption{ Distance distributions of RR Lyrae stars (left) and Cepheids (right) in our \texttt{eRASS1} cross-matched sample, using distances from \textit{Gaia} \citep{2021AJ....161..147B}. Blue histograms show the full pulsator samples, while vertical dashed red lines indicate those with candidate X-ray counterparts. }
         \label{fig:dist_hist}
   \end{figure*}

To calculate the X-ray luminosities, we adopted pulsator distances derived from \textit{Gaia} parallaxes \citep{2021AJ....161..147B}.  As shown in Fig.~\ref{fig:luminosity_errorbars}, many of the detected RR~Lyrae stars appear to have X-ray luminosities comparable to the brightest Cepheids. This is unexpected, given that no RR~Lyrae star has been previously confirmed as an X-ray source, while a few Cepheids have known detections, despite RR~Lyrae stars being far more numerous in the Milky Way. Moreover, the RR~Lyrae stars with apparent X-ray counterparts exhibit a nearly uniform distribution in distance, suggesting that most detections are likely the result of spurious associations with nearby background sources or rare RR~Lyrae binary companions (Fig.~\ref{fig:dist_hist}). In contrast, the distribution of detected Cepheids follows the expected inverse-square law trend, consistent with genuine X-ray detections.

Taken together, these findings indicate that the majority of RR~Lyrae X-ray associations are likely due to chance alignments or unresolved companions, whereas the Cepheid detections are more reliable. The most plausible RR~Lyrae X-ray detection in our sample is \object{OGLE-BLG-RRLYR-00252}, which is among the closest RR~Lyrae stars in our sample. If this detection is genuine and represents the upper bound of RR~Lyrae X-ray luminosity, it implies that RR~Lyrae stars are intrinsically at least two orders of magnitude fainter in X-rays compared to the most luminous Cepheids.

\begin{figure}
   \centering
   \includegraphics[width=0.99\columnwidth]{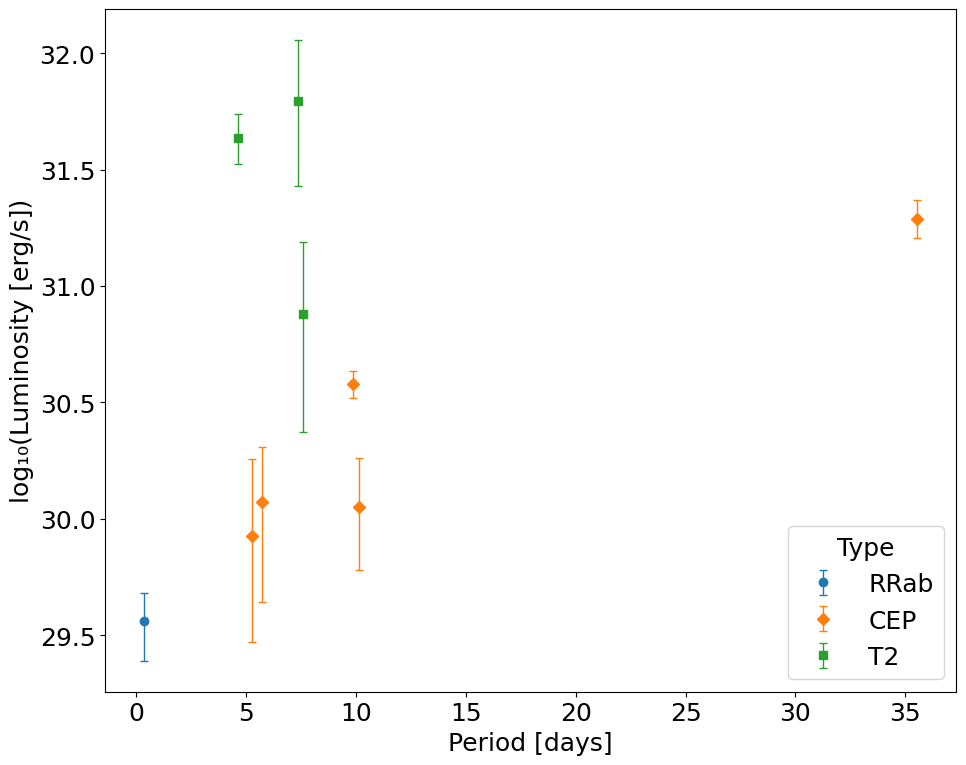}
      \caption{X-ray luminosity as a function of pulsation period for classically pulsating stars with X-ray counterparts detected in \texttt{eRASS1}. Tentative X-ray associations for RR Lyrae stars have been excluded for clarity.}
         \label{fig:P-lum}
   \end{figure}

The relationship between pulsation period and X-ray luminosity is shown in Fig.~\ref{fig:P-lum}. The data suggest a possible trend of increasing X-ray luminosity with pulsation period among DCEPs, although this result should be interpreted with caution, as some of these stars are affected by optical loading. Notably, type II Cepheids appear more X-ray-luminous than DCEPs at comparable pulsation periods (approximately 5 to 10 days). This may reflect intrinsic structural differences between the two classes and could offer insights into the mechanisms driving X-ray production in classical pulsators. However, we note that the short average exposure time of \texttt{eRASS1} \citep[170~seconds;][]{2024A&A...681A..77Z} means that each pulsating star is typically observed at a single, random phase of its pulsation cycle. If the X-ray emission varies significantly throughout the cycle, as has been observed in some Cepheids, this could introduce scatter in the measured fluxes and bias detections towards phases of higher luminosity.

\section{Discussion}

The detection of X-ray emission in classical pulsators offers a valuable diagnostic for probing atmospheric dynamics driven by stellar pulsation. In DCEPs, previous observations with \textit{XMM-Newton} and \textit{Chandra} \citep[e.g.][]{engle2017, evans2020_xray_v473lyr} have demonstrated that their X-ray flux varies with pulsation phase. This modulation strongly suggests a connection to pulsation-driven shock waves propagating through the stellar atmosphere. These shocks, generated by high-amplitude radial pulsations, compress and heat atmospheric layers, potentially producing X-rays through both thermal bremsstrahlung and non-thermal mechanisms \citep{moschou2020, Fraschetti2023}.

Theoretical predictions have long suggested that RR Lyrae stars should also be soft X-ray sources, powered by similar shock processes \citep{bejgman.stepanov1981, andrievskii1993}. Yet, observational confirmation has remained elusive. The tentative detection of X-ray emission from OGLE-BLG-RRLYR-00252 in this study may represent the first empirical support for these predictions. Interestingly, although RR Lyrae stars exhibit stronger pulsational shocks that are often hypersonic, with Mach numbers exceeding 10 \citep{Gillet2019}, they appear intrinsically fainter in X-rays compared to Cepheids, where shocks are typically less intense and reach only moderate Mach numbers \citep{Gillet2014}. 

A tentative linear correlation between X-ray luminosity and pulsation period among DCEPs is evident in Fig.~\ref{fig:P-lum}. This trend suggests that the stellar radius directly influences X-ray brightness. Larger stars may possess more extended atmospheres, supporting bigger or more energetic shock fronts that enhance the high-energy emission. Our most promising RR Lyrae candidate, OGLE-BLG-RRLYR-00252, appears to lie along a short-period extension of the relation defined by DCEPs, implying a possible continuity in the shock-driven X-ray mechanism across pulsator classes. Type II Cepheids, by contrast, appear more X-ray-luminous than DCEPs at similar periods and do not follow the same linear trend. This deviation may indicate that other processes, such as enhanced mass loss, wind-driven shocks, or structural differences associated with post-asymptotic giant branch evolution, contribute to the higher X-ray output observed in Type II Cepheids.  However, our candidate detections should be independently confirmed and their X-ray luminosity levels should be measured before more concrete conclusions can be drawn.

     \begin{figure*}[h]
   \centering
   \includegraphics[width=0.99\columnwidth]{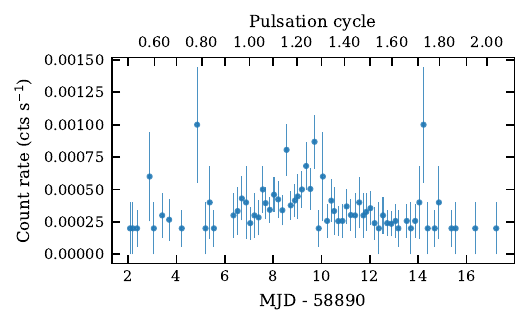}
   \includegraphics[width=0.99\columnwidth]{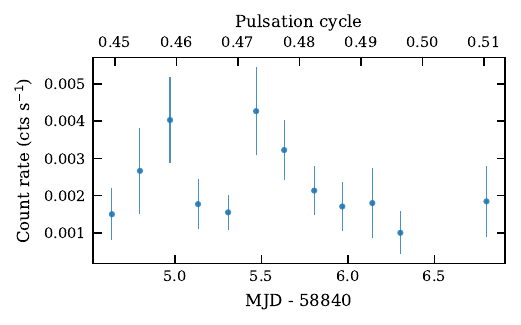}
     \caption{eROSITA light curves in the 0.2--10~keV band for bet~Dor (left) and l~Car (right). Pulsation phase 0.0 corresponds to the maximum of the optical brightness.}
   \label{fig:xray}
   \end{figure*}

Some X-ray detections from pulsators may originate not from the pulsating stars themselves, but from unresolved binary companions. In main-sequence stars, X-ray emission is closely tied to rotation-driven magnetic activity, which declines with age as stellar rotation slows. Although classical pulsators have undergone multiple evolutionary stages and their present-day rotation rates no longer follow standard magnetic–age relations, DCEPs are relatively young, whereas RR~Lyrae stars are among the oldest stellar populations in the Galaxy. For the pulsators themselves, it is therefore unclear whether age plays a direct role in their X-ray output, since their emission is more likely produced by pulsation-driven shocks than by rotation-generated magnetism. However, age becomes important when considering unresolved companions. Young low-mass stars commonly exhibit strong magnetic activity and higher X-ray luminosities, while old low-mass stars are typically X-ray faint. Consequently, companions to DCEPs would be expected to be relatively young and potentially X-ray bright, increasing the likelihood of companion contamination. In contrast, any companions to RR~Lyrae stars would generally be old and magnetically inactive, making significant X-ray contamination far less likely \citep{2021ApJ...915...50H}. This distinction further supports the interpretation that most RR~Lyrae associations in our sample are spurious rather than companion-driven, while companion contributions remain a viable explanation for some Cepheid detections.

Several Cepheids in our sample are known or suspected to host companions, making contamination a plausible scenario. Among the DCEPs with X-ray counterparts, AX~Cir and MY~Pup are particularly relevant. AX~Cir is a single-lined spectroscopic binary with an orbital period exceeding 6000\,days \citep{veloce2}, and is likely part of a hierarchical multiple system \citep{2019A&A...623A.117K}, with indications that the secondary may itself be a binary. Such a configuration raises the possibility that one or more low-mass companions could be magnetically active and X-ray-luminous. MY~Pup also appears to host a companion cooler than B9 \citep{1992ApJ...384..220E}, again making a low-mass, active companion a plausible source of the observed X-rays. In both systems, companion-generated emission could contribute to, or even dominate, the measured flux, and cannot be excluded without further observational constraints.

Whether the X-ray emission originates from the Cepheid or from a companion can in principle be determined through time-resolved observations. For example, $\delta$~Cep has been shown to exhibit phase-dependent X-ray variability, confirming that the emission arises from the Cepheid itself despite the presence of a companion \citep{engle2017}. Such variability therefore provides a key diagnostic for identifying the true source of the X-rays.

To investigate the temporal variability of the X-ray sources detected in \texttt{eRASS1}, we extracted light curves in the 0.2--10~keV band using the eROSITA Science Analysis Software System (\texttt{eSASS}; \citealt{2022A&A...661A...1B}) with 1000\,s time bins. For most objects, the low signal-to-noise ratio precluded firm conclusions about variability, as the uncertainties within individual bins often dominated the signal. Nevertheless, we provide a comparison of the X-ray light curves with the optical data in Appendix~\ref{app:phaseplots}, since the observations were frequently obtained over a narrow range of pulsation phases. This comparison offers at least a limited constraint on the behaviour as a function of phase.

Clear temporal variability in X-rays was detected only for the DCEPs bet~Dor and l~Car. Although our observations did not span multiple pulsation cycles, and in the case of l~Car did not even cover a full cycle, both stars showed likely phase-dependent changes (Fig.~\ref{fig:xray}). Such behaviour was already observed for bet~Dor \citep{2009AIPC.1135..192E}, but is detected for the first time in l~Car \citep{2022ApJ...938..153E}. This favours intrinsic rather than companion-driven X-ray emission, although in the case of l~Car confirmation will require observations that extend over several full cycles. Both Cepheids also showed what appear to be multiple peaks in their X-ray light curves, which, if confirmed, would suggest that X-ray production in these stars is even more complex than previously assumed. Because we could not study the variation of X-ray luminosity with pulsation phase in the remaining objects, future phase-resolved observations will be crucial for determining the true origin of the high-energy radiation in candidate X-ray-emitting type~II Cepheids and RR~Lyrae stars.

\section{Conclusions}\label{sec:conclusions}

We have conducted a systematic cross-match between RR~Lyrae and Cepheid stars and X-ray sources in the \texttt{eRASS1} catalogue, identifying seven RR~Lyrae and eight Cepheid candidates with potential X-ray counterparts. While most Cepheid associations appear reliable and include both classical and type II pulsators, the RR~Lyrae associations are dominated by likely misidentifications, spurious alignments, or unresolved binaries.

Although RR~Lyrae stars experience stronger pulsational shocks than Cepheids --- due to their shorter periods and larger radial velocity amplitudes --- they are intrinsically fainter in X-rays. This is consistent with previous studies, which suggest that RR~Lyrae stars are weak X-ray emitters. Among our RR~Lyrae sample, only one source, OGLE-BLG-RRLYR-00252, emerges as a plausible genuine detection. If confirmed, its X-ray luminosity would set an upper limit for RR~Lyrae stars at approximately two orders of magnitude below that of the most X-ray-luminous Cepheids.

We observe a tentative correlation between X-ray luminosity and pulsation period among DCEPs, as well as systematically higher X-ray brightness in type~II Cepheids at comparable periods. While these trends may reflect intrinsic differences in pulsation-driven activity, they must be interpreted with caution due to the short exposure times and the potential phase dependence of the X-ray emission. If confirmed, this pronounced difference would offer a valuable opportunity to probe the underlying physics of shock propagation and energy dissipation in stellar atmospheres. In particular, it may be linked to differences in atmospheric structure and shock efficiency among these stars.

Targeted follow-up studies, particularly with \textit{XMM-Newton} or \textit{Chandra}, will be essential for confirming these detections and constraining the physical mechanisms responsible for X-ray emission in classical pulsators. Phase-resolved observations would allow direct verification of the pulsating star as the X-ray source and enable detailed comparisons of shock dynamics and atmospheric structure in RR~Lyrae stars and Cepheids. Such observations would provide critical insights into the conditions that facilitate X-ray generation in these variable stars.

\begin{acknowledgements}
KI is supported by the Polish National Science Centre (NCN) grant 2024/55/D/ST9/01713.
HN is funded by the European Research Council (ERC) under the European Union’s Horizon 2020 research and innovation programme (grant agreement No. 951549 - UniverScale). 

This work is based on data from eROSITA, the soft X-ray instrument aboard SRG, a joint Russian-German science mission supported by the Russian Space Agency (Roskosmos), in the interests of the Russian Academy of Sciences represented by its Space Research Institute (IKI), and the Deutsches Zentrum für Luft- und Raumfahrt (DLR). The SRG spacecraft was built by Lavochkin Association (NPOL) and its subcontractors, and is operated by NPOL with support from the Max Planck Institute for Extraterrestrial Physics (MPE). The development and construction of the eROSITA X-ray instrument was led by MPE, with contributions from the Dr. Karl Remeis Observatory Bamberg \& ECAP (FAU Erlangen-Nuernberg), the University of Hamburg Observatory, the Leibniz Institute for Astrophysics Potsdam (AIP), and the Institute for Astronomy and Astrophysics of the University of Tübingen, with the support of DLR and the Max Planck Society. The Argelander Institute for Astronomy of the University of Bonn and the Ludwig Maximilians Universität Munich also participated in the science preparation for eROSITA.

The eROSITA data shown here were processed using the eSASS software system developed by the German eROSITA consortium.

We acknowledge with thanks the variable star observations from the AAVSO International Database contributed by observers worldwide and used in this research."
\end{acknowledgements}

\bibpunct{(}{)}{;}{a}{}{,} 
\bibliographystyle{aa} 
\bibliography{references}

\begin{appendix}

\onecolumn
\section{Phase plots of X-ray and optical data}\label{app:phaseplots}

\begin{figure*}[h]
   \centering
   \includegraphics[width=0.3\textwidth]{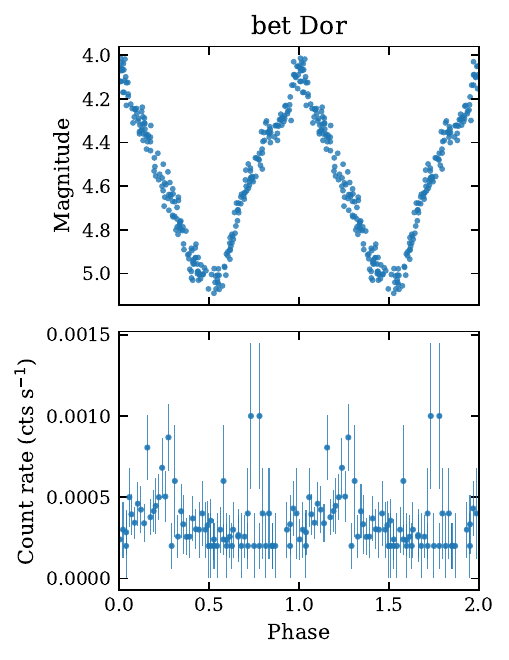}
   \includegraphics[width=0.3\textwidth]{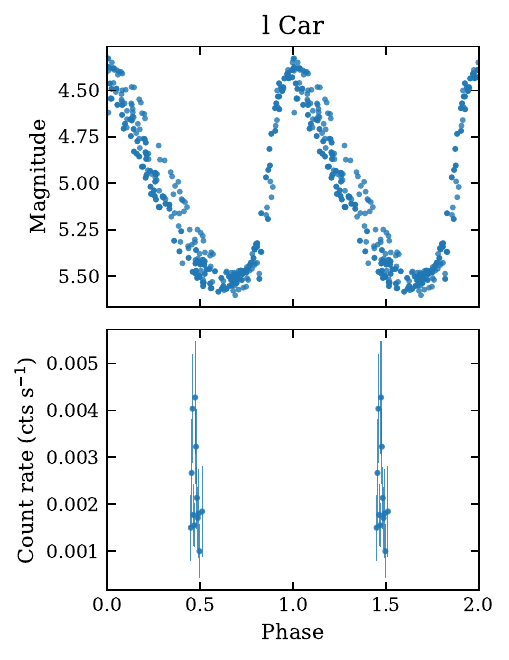}
   \includegraphics[width=0.3\textwidth]{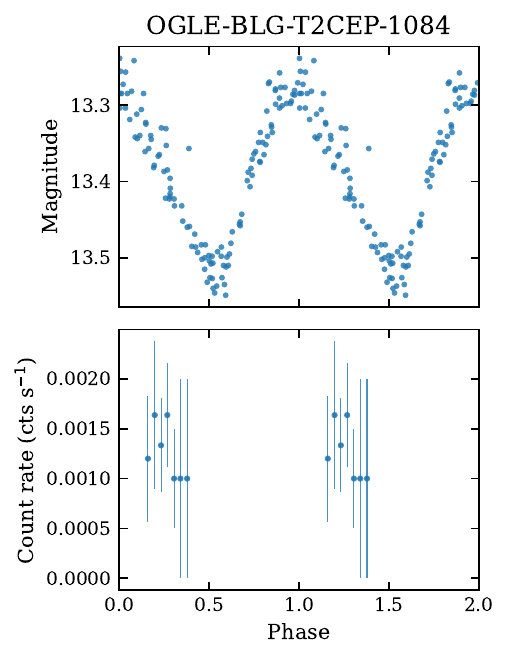}
   
   \includegraphics[width=0.3\textwidth]{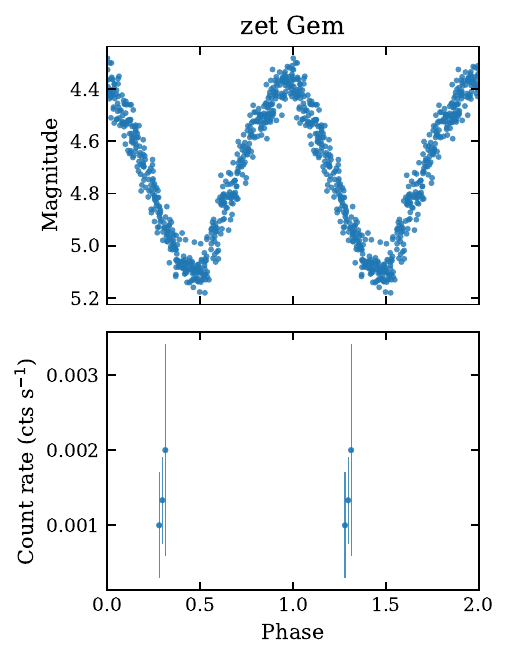}
   \includegraphics[width=0.3\textwidth]{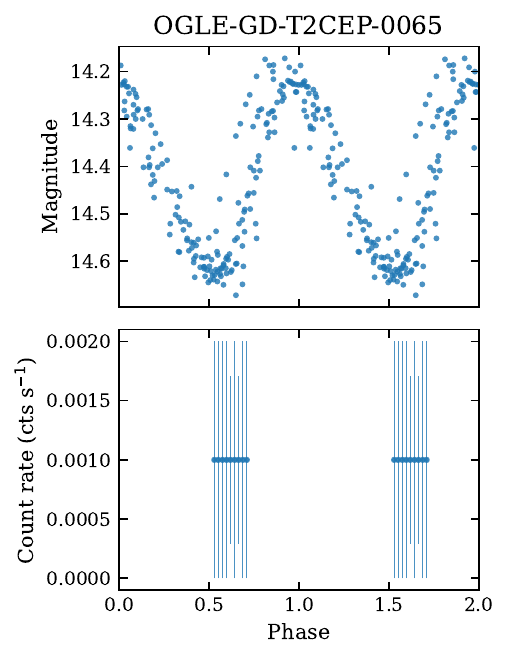}
   \includegraphics[width=0.3\textwidth]{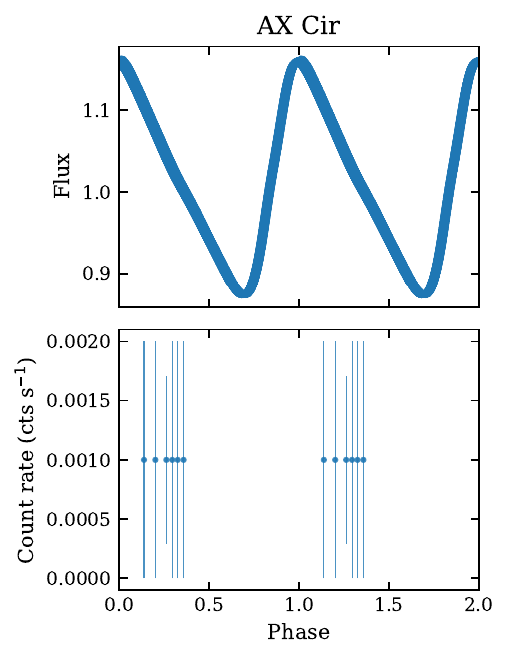}
   
   \includegraphics[width=0.3\textwidth]{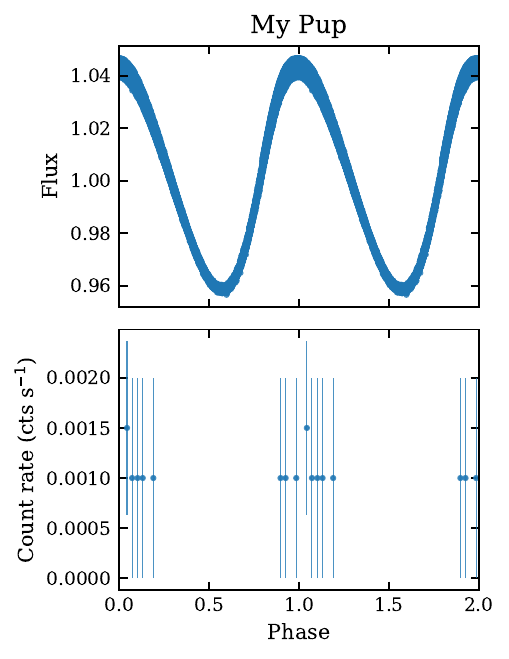}
   \includegraphics[width=0.3\textwidth]{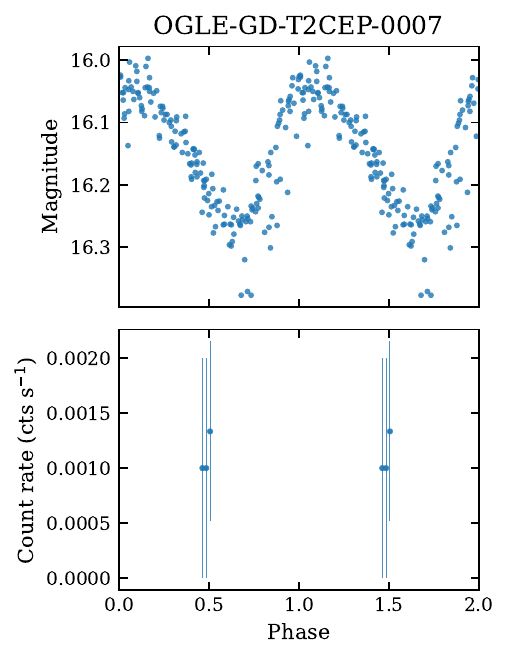}
   \includegraphics[width=0.3\textwidth]{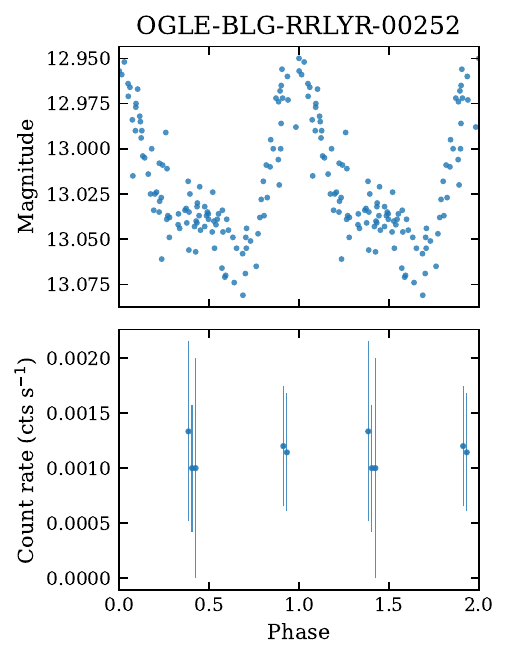}
   
      \caption{Comparison of optical and X-ray light curves phased with the pulsation period. The optical data are from the AAVSO database (bet~Dor, l~Car, and zet~Gem), the TESS satellite (AX~Cir and MY~Pup), and the OGLE Survey (remaining objects).    }
         \label{fig:phaseplots}
   \end{figure*}

The X-ray and optical data phased with pulsation period is presented in Fig.~\ref{fig:phaseplots}.

\twocolumn

\end{appendix}
\end{document}